\documentclass[aps,prl,twoside,twocolumn,floatfix,amsmath,showpacs,amssymb,nofootinbib]{revtex4}

 \usepackage[latin1]{inputenc}
 \usepackage{amsmath}
 \usepackage{subfigure}
 \usepackage{placeins}
 \usepackage{floatflt} \usepackage{amsmath}
 \usepackage{amssymb}
 \usepackage{wrapfig}
 \usepackage[]{tocbibind}
 \usepackage[all]{xy}

 \usepackage{ifthen}
 \usepackage{times}
 
\makeatletter
\newcommand\figcaption{\def\@captype{figure}\caption}
\newcommand\tabcaption{\def\@captype{table}\caption}
\makeatother

\usepackage[pdftex]{graphicx}                                

\begin{document}
\DeclareGraphicsExtensions{.jpg,.pdf,.png,.mps,.eps,.ps}  

\title{Subthreshold $\mathbf{\Xi^-}$ Production in Collisions of p\,(3.5~GeV)\,+\,Nb}

\author{
G.~Agakishiev$^{7}$, O.~Arnold$^{9}$, A.~Balanda$^{3}$, D.~Belver$^{18}$, A.\,V.~Belyaev$^{7}$, 
J.\,C.~Berger-Chen$^{9}$, A.~Blanco$^{2}$, M.~B\"{o}hmer$^{10}$, J.\,L.~Boyard$^{16}$, 
P.~Cabanelas$^{18,a}$, S.~Chernenko$^{7}$, A.~Dybczak$^{3}$, E.~Epple$^{9}$, L.~Fabbietti$^{9}$, 
O.\,V.~Fateev$^{7}$, P.~Finocchiaro$^{1}$, P.~Fonte$^{2,b}$, J.~Friese$^{10}$, 
I.~Fr\"{o}hlich$^{8}$, T.~Galatyuk$^{5,c}$, J.\,A.~Garz\'{o}n$^{18}$, R.~Gernh\"{a}user$^{10}$, 
K.~G\"{o}bel$^{8}$, M.~Golubeva$^{13}$, D.~Gonz\'{a}lez-D\'{\i}az$^{5}$, F.~Guber$^{13}$, 
M.~Gumberidze$^{5,16}$, T.~Heinz$^{4}$, T.~Hennino$^{16}$, R.~Holzmann$^{4}$, 
A.~Ierusalimov$^{7}$, I.~Iori$^{12,d}$, A.~Ivashkin$^{13}$, M.~Jurkovic$^{10}$, 
B.~K\"{a}mpfer$^{6,e}$, T.~Karavicheva$^{13}$, I.~Koenig$^{4}$, W.~Koenig$^{4}$, 
B.\,W.~Kolb$^{4}$, G.~Kornakov$^{18}$, 
R.~Kotte$^{6}$, 
A.~Kr\'{a}sa$^{17}$, F.~Krizek$^{17}$, 
R.~Kr\"{u}cken$^{10}$, H.~Kuc$^{3,16}$, W.~K\"{u}hn$^{11}$, A.~Kugler$^{17}$, A.~Kurepin$^{13}$, 
V.~Ladygin$^{7}$, R.~Lalik$^{9}$, S.~Lang$^{4}$, K.~Lapidus$^{9}$, A.~Lebedev$^{14}$, 
T.~Liu$^{16}$, L.~Lopes$^{2}$, M.~Lorenz$^{8,c}$, L.~Maier$^{10}$, A.~Mangiarotti$^{2}$, 
J.~Markert$^{8}$, V.~Metag$^{11}$, B.~Michalska$^{3}$, J.~Michel$^{8}$, C.~M\"{u}ntz$^{7}$, 
L.~Naumann$^{6}$, Y.\,C.~Pachmayer$^{8}$, M.~Palka$^{3}$, Y.~Parpottas$^{15,f}$, 
V.~Pechenov$^{4}$, O.~Pechenova$^{8}$, J.~Pietraszko$^{4}$, W.~Przygoda$^{3}$, 
B.~Ramstein$^{16}$, A.~Reshetin$^{13}$, A.~Rustamov$^{8}$, A.~Sadovsky$^{13}$, 
P.~Salabura$^{3}$, A.~Schmah$^{9,g}$, E.~Schwab$^{4}$, J.~Siebenson$^{9}$, Yu.\,G.~Sobolev$^{17}$,
S.~Spataro$^{11,h}$, B.~Spruck$^{11}$, H.~Str\"{o}bele$^{8}$, J.~Stroth$^{8,4}$, C.~Sturm$^{4}$, 
A.~Tarantola$^{8}$, K.~Teilab$^{8}$, P.~Tlusty$^{17}$, M.~Traxler$^{4}$, R.~Trebacz$^{3}$, 
H.~Tsertos$^{15}$, T.~Vasiliev$^{7}$, V.~Wagner$^{17}$, M.~Weber$^{10}$, 
C.~Wendisch$^{4}$, 
J.~W\"{u}stenfeld$^{6}$, S.~Yurevich$^{4}$, Y.\,V.~Zanevsky$^{7}$ \\[10bp]
(HADES collaboration)\vspace*{10bp}
}
\affiliation{
\mbox{ $^{1}$~Instituto Nazionale di Fisica Nucleare - Laboratori Nazionali del Sud,
95125~Catania, Italy}\\
\mbox{ $^{2}$~LIP-Laborat\'{o}rio de Instrumenta\c{c}\~{a}o e F\'{\i}sica
Experimental de Part\'{\i}culas , 3004-516~Coimbra, Portugal}\\
\mbox{ $^{3}$~Smoluchowski Institute of Physics, Jagiellonian University of Cracow,
30-059~Krak\'{o}w, Poland}\\
\mbox{ $^{4}$~GSI Helmholtzzentrum f\"{u}r Schwerionenforschung GmbH,
64291~Darmstadt, Germany}\\
\mbox{ $^{5}$~Technische Universit\"{a}t Darmstadt, 64289~Darmstadt, Germany}\\
\mbox{ $^{6}$~Institut f\"{u}r Strahlenphysik, Helmholtz-Zentrum Dresden-Rossendorf,
01328~Dresden, Germany}\\
\mbox{ $^{7}$~Joint Institute of Nuclear Research, 141980~Dubna, Russia}\\
\mbox{ $^{8}$~Institut f\"{u}r Kernphysik, Johann Wolfgang Goethe-Universit\"{a}t,
60438 ~Frankfurt, Germany}\\
\mbox{ $^{9}$~Excellence Cluster 'Origin and Structure of the Universe', 
85748~Garching, Germany}\\
\mbox{$^{10}$~Physik Department E12, Technische Universit\"{a}t M\"{u}nchen,
85748~Garching, Germany}\\
\mbox{$^{11}$~II.Physikalisches Institut, Justus Liebig Universit\"{a}t Giessen,
35392~Giessen, Germany}\\
\mbox{$^{12}$~Istituto Nazionale di Fisica Nucleare, Sezione di Milano,
20133~Milano, Italy}\\
\mbox{$^{13}$~Institute for Nuclear Research, Russian Academy of Science,
117312~Moscow, Russia}\\
\mbox{$^{14}$~Institute of Theoretical and Experimental Physics, 117218~Moscow, Russia}\\
\mbox{$^{15}$~Department of Physics, University of Cyprus, 1678~Nicosia, Cyprus}\\
\mbox{$^{16}$~Institut de Physique Nucl\'{e}aire (UMR 8608),
CNRS/IN2P3 - Universit\'{e} Paris Sud, F-91406~Orsay Cedex, France}\\
\mbox{$^{17}$~Nuclear Physics Institute, Academy of Sciences of Czech Republic, 
25068~Rez, Czech Republic}\\
\mbox{$^{18}$~LabCAF F. F\'{i}sica, Univ. de Santiago de Compostela, 15706~Santiago de Compostela, Spain}\\ 
\\
\mbox{ $^{a}$~also at Nuclear Physics Center of University of Lisbon, 1649-013 Lisboa, Portugal}\\
\mbox{ $^{b}$~also at ISEC Coimbra, 3030-199 Coimbra, Portugal}\\
\mbox{ $^{c}$~also at ExtreMe Matter Institute EMMI, 64291~Darmstadt, Germany}\\
\mbox{ $^{d}$~also at Dipartimento di Fisica, Universit\`{a} di Milano, 20133 Milano, Italy}\\
\mbox{ $^{e}$~also at Technische Universit\"{a}t Dresden, 01062~Dresden, Germany}\\
\mbox{ $^{f}$~also at Frederick University, 1036~Nikosia, Cyprus}\\
\mbox{ $^{g}$~now at Lawrence Berkeley National Laboratory, Berkeley, USA}\\
\mbox{ $^{h}$~now at Dipartimento di Fisica Generale and INFN, Universit\`{a} di Torino, 10125 Torino, Italy}\\ 
}
\date{\today}

\begin{abstract}
Results on the production of the double-strange 
cascade hyperon $\mathrm{\Xi^-}$ are reported for collisions of p\,(3.5~GeV)\,+\,Nb, 
studied with the High Acceptance Di-Electron Spectrometer (HADES) at SIS18 at GSI 
Helmholtzzentrum for Heavy-Ion Research, Darmstadt. For the first time, subthreshold 
$\mathrm{\Xi^-}$ production is observed in proton-nucleus interactions. 
Assuming a $\mathrm{\Xi^-}$ phase-space distribution similar to that of $\mathrm{\Lambda}$ hyperons, 
the production probability amounts to  
$P_{\mathrm{\Xi^-}}=(2.0\,\pm0.4\,\mathrm{(stat)}\,\pm 0.3\,\mathrm{(norm)}\,\pm 0.6\,\mathrm{(syst)})\times10^{-4}$ 
resulting in a $\mathrm{\Xi^-/(\Lambda+\Sigma^0)}$ ratio of 
$P_{\mathrm{\Xi^-}}/\ P_{\mathrm{\Lambda+\Sigma^0}}=(1.2\pm 0.3\,\mathrm{(stat)}\pm0.4\,\mathrm{(syst)})\times10^{-2}$. 
Available model predictions are significantly lower than the estimated $\mathrm{\Xi^-}$ yield.
\end{abstract}


\maketitle
The double-strange $\mathrm{\Xi^-}$ baryon (also known as cascade particle) 
when produced in elementary nucleon-nucleon (NN) collisions  
must be co-produced with two kaons ensuring strangeness conservation, 
$\mathrm{NN \rightarrow N\Xi KK}$. In fixed-target experiments, this requires a minimum beam energy
of $E_{thr}=3.74$~GeV ($\sqrt{s_{thr}}=3.25$~GeV). 
In heavy-ion and even in nucleon-nucleus collisions cooperative processes are possible 
allowing for the production below this threshold. Above threshold and in 
heavy-ion reactions, the $\mathrm{\Xi^-}$ hyperons were measured over about three orders of 
magnitude of the centre-of-mass energy covered by the 
LHC ($\sqrt{s_{NN}}=2.76$~TeV \cite{ALICE_Xi_PbPb_2p76TeV}), 
RHIC ($\sqrt{s_{NN}}=62.4, 200$~GeV \cite{STAR11,STAR07}), 
SPS ($\sqrt{s_{NN}}=8.9, 17.3$~GeV \cite{NA57_04}, $\sqrt{s_{NN}}=6.4-17.3$~GeV) \cite{NA49_08}), 
and AGS ($\sqrt{s_{NN}}=3.84$~GeV \cite{E895_04}) accelerators.  
The yield of multi-strange particles produced 
in nucleon-nucleus (p\,+\,$A$) and nucleus-nucleus ($A$\,+\,$A$)
collisions below their production 
threshold in NN collisions, is expected to be sensitive to the equation of state (EoS) 
of nuclear matter, similar to single-strange hadrons \cite{Hartnack03,Fuchs06,Hartnack12}.
In heavy-ion reactions, the necessary energy for the production of multi-strange hyperons can be  
accumulated via multiple collisions involving nucleons, produced particles and short-living 
resonances. The corresponding number of such collisions increases with the density within the 
reaction zone the maximum of which in turn depends on the stiffness of the EoS. 

Predictions of sub-threshold cascade production  
at energies available with the heavy-ion synchrotron SIS18 at GSI, Darmstadt,  
were made within a relativistic transport model 
\cite{CheMingKo04}. The cross sections of the strangeness exchange reactions 
$\mathrm{\overline{K} Y \rightarrow \pi \Xi}$ ($\mathrm{Y=\Lambda,\,\Sigma}$), 
which were thought to be essential for $\mathrm{\Xi}$ creation below NN threshold, were 
taken from a coupled-channel approach based on a flavor SU(3)-invariant hadronic Lagrangian 
\cite{CheMingKo02}.  
At that time, no subthreshold $\mathrm{\Xi^-}$ production was observed; the first  
announcement came from the HADES collaboration studying, at SIS18, Ar\,+\,KCl reactions 
at a beam kinetic energy of 1.76$A$~GeV ($\sqrt{s_{NN}}=2.61$~GeV \cite{hades_Xi_ArKCl}). 
The deduced $\mathrm{\Xi^-/\Lambda}$ ratio was found 
substantially larger than any model prediction available at that time. Shortly after, 
other strangeness-exchange reactions, e.g. the hyperon-hyperon scattering processes, 
$\mathrm{YY \rightarrow \Xi N}$, exhibiting quite high cross sections, were figured out 
to largely account for this discrepancy \cite{FengLi12,Kolomeitsev12}, while the reaction 
$\mathrm{\overline{K} Y \rightarrow \pi \Xi}$ was found negligible \cite{FengLi12}. Also, 
a very recent investigation \cite{Graef14} of deep-subthreshold $\mathrm{\Xi}$ production 
in nuclear collisions with the UrQMD transport model \cite{UrQMD1,UrQMD2} making use of 
the $\mathrm{YY}$ cross sections provided in ref.~\cite{FengLi12} 
showed that the hyperon strangeness exchange is the 
dominant process  contributing to the $\mathrm{\Xi}$ yield. However, the model could not 
satisfactorily explain the $\mathrm{\Xi^-/\Lambda}$ ratio measured by HADES.
Presently, no experimental data exist on $\mathrm{\Xi}$ production in p\,+\,$A$ interactions 
near threshold. The lowest energy so far at which $\mathrm{\Xi^-}$ production has been 
observed in collisions of p\,+\,Be and p\,+\,Pb is the maximum SPS energy 
($\sqrt{s_{NN}}=17.3$~GeV \cite{NA57_06}). 

It would be interesting to learn which processes contribute mainly to 
subthreshold $\mathrm{\Xi}$ production in case of nucleon-nucleus collisions, which are 
considered as a link between elementary NN and heavy-ion collisions. 
For instance, in p\,+\,$A$ reactions at a beam kinetic energy of 3.5~GeV 
(fixed target, $\sqrt{s_{NN}}=3.18$~GeV), 
the scattering of two incoherently produced hyperons appears 
rather improbable. Also, at first sight, direct double-hyperon production seems to be 
impossible, since the threshold of $\sqrt{s_{thr, \mathrm{\Lambda\Lambda}}}=3.22$~GeV 
for the channel requiring the lowest energy effort, 
$\mathrm{p p \rightarrow \Lambda \Lambda K^+ K^+}$, is only marginally lower than the 
$\mathrm{\Xi}$ threshold. However, already the consideration of a rather modest Fermi 
motion of the nucleons within the nucleus, i.e. a counter-motion against the 
projectile with a momentum of about 50~MeV$/c$ (being well below the Fermi momentum), 
would lift the available energy above both thresholds. 
Another possibility to gain the necessary energy would be the scattering of the proton 
at an object acting more massively than a single nucleon in the nucleus. 
Thus, for p\,(3.5~GeV)\,+\,$A$ collisions, a target object $\mathrm{X}$ as heavy as only 1.11 
nucleon masses is sufficient to reach the threshold energy of the final state 
$\mathrm{X \Xi K K}$. Hence, the co-operation of two, or more, correlated target nucleons, 
e.g. bound in $\alpha$ particles, would allow for this kinematic effect.     
Correspondingly, a high collectivity of the target nucleus was already observed in 
deep-subthreshold kaon production in p\,+\,$A$ ($A\,$= C, Cu, Au) collisions 
at a beam kinetic energy of 1.0~GeV studied by ANKE at COSY-J\"{u}lich \cite{ANKE01}. 
Also, in electron scattering experiments, $^{12}$C(e,e'p) at 4.627~GeV, performed     
at JLab \cite{Subedi08}, a surprisingly high fraction of nucleons, i.e. 20\,\%, are found 
to be strongly correlated, predominantly in the form of proton-neutron pairs.
Finally, $\mathrm{\Xi}$ production in p\,+\,$A$ (and $A$\,+\,$A$) might happen via an exotic 
channel as $\mathrm{\eta\Lambda\rightarrow \Xi K}$, probably via the excitation and decay of 
a massive resonance as, e.g., $\mathrm{\Lambda^*(2000, 2100, ...)}$ \cite{PDG2014}. 
Another possible scenario is to produce a non-strange heavy resonance with a mass 
sufficiently high to decay into $\mathrm{\Xi K K}$. This hypothesis 
is motivated by the fact that in p\,+\,p collisions at 3.5\,GeV beam kinetic energy a 
substantial fraction of the exclusive production of 
$\mathrm{\Sigma(1385)^+ + K^+ + n}$ was found to proceed via an intermediate broad 
$\mathrm{\Delta^{++}}$ excitation at about 2000\,MeV \cite{hades_Sigma1385}.  

In this Letter, we report on the first observation of sub-threshold $\mathrm{\Xi^-}$ 
production in nucleon-nucleus collisions at $\sqrt{s_{NN}}-\sqrt{s_{thr}}=-70$\,MeV. 
The experiment was performed 
with the {\bf H}igh {\bf A}cceptance {\bf D}i-{\bf E}lectron {\bf S}pectrometer (HADES)
at the Schwerionensynchrotron SIS18 at GSI, Darmstadt. HADES, although
primarily optimized to measure di-electrons \cite{HADES-PRL07}, offers also excellent
hadron identification capabilities
\cite{hades_K0_ArKCl,hades_Lambda_ArKCl,hades_kpm_phi}.
A detailed description of the spectrometer is presented in ref.~\cite{hades_spectro}.
The present results are based on a dataset which was previously
analyzed with respect to e$^+$e$^-$ \cite{hades_epm_pNb} as well as to pion and 
$\mathrm{\eta}$ \cite{hades_pion_eta_pNb}, K$^0$ \cite{hades_pp_pNb_K0} and 
$\mathrm{\Lambda}$ \cite{hades_lambda_pNb} production in collisions of p\,+\,Nb at 
3.5\,GeV. The main features of the apparatus relevant for the present analysis 
are summarized in ref.~\cite{hades_lambda_pNb}. 

In the present experiment, a proton beam of about $2\times 10^6$ particles per second 
with kinetic energy of 3.5~GeV was incident on a 12-fold segmented target of natural 
niobium ($^{93}$Nb). The data readout was started by different trigger decisions 
\cite{hades_pion_eta_pNb}. For the present analysis, 
we employ only the data of the first-level (LVL1) trigger, 
requiring a charged-particle multiplicity $\ge 3$ in the time-of-flight wall 
composed of plastic scintillation detectors. 
We processed about $N_{\mathrm{LVL1}}=3.2\times10^{9}$ of such LVL1 events. 

It is important to mention that $\mathrm{\Sigma^0}$ hyperons decay almost exclusively 
into $\mathrm{\Lambda}$'s via the decay $\mathrm{\Sigma^0 \rightarrow \Lambda \gamma}$ 
(branching ratio $BR=100$\,\%, $c\tau=2.22\times10^{-11}$\,m \cite{PDG2014}), with  
the photon not being detected in the present experiment. 
Hence, throughout the paper, any ``$\mathrm{\Lambda}$ yield''   
has to be understood as that of $\mathrm{\Lambda+\Sigma^0}$. Correspondingly, in case of  
simulations, where the individual particle species are known, 
the yields of $\mathrm{\Lambda}$ and $\mathrm{\Sigma^0}$ hyperons are summed up. 

In the present analysis, we identify the $\mathrm{\Xi^-}$ and $\mathrm{\Lambda}$ hyperons 
through their weak decays $\mathrm{\Xi^- \rightarrow \Lambda \pi^-}$ ($BR=99.9$\,\%, 
$c\tau=4.91$\,cm) and $\mathrm{\Lambda \rightarrow p \pi^-}$ ($BR=63.9$\,\%, 
$c\tau=7.89$\,cm) \cite{PDG2014}, with the charged hadrons detected in 
HADES \cite{hades_Xi_ArKCl,hades_Lambda_ArKCl,hades_lambda_pNb}. 
The long lifetimes cause a sizeable fraction   
of these particles to decay away from the primary vertex. The precision of the track
reconstruction with HADES is sufficient to resolve these secondary vertices 
\cite{hades_Xi_ArKCl}. To allow for $\mathrm{\Lambda}$ selection various topological 
cuts on single-particle and two-particle quantities were applied. These are 
i) a minimum value of the proton track\footnote{With ``track'' we mean the trajectory 
of a particle track extrapolated up to the relevant vertex.}
 distance to the primary vertex (\mbox{p-VecToPrimVer}), 
ii) the same for the $\pi^-$ (\mbox{$\pi_1$-VecToPrimVer}), 
iii) an upper limit of the p-$\pi^-$ minimum track distance (\mbox{p-$\pi_1$-MinVecDist}), 
and iv) a minimum value of the $\mathrm{\Lambda}$ decay vertex distance to the primary vertex 
(\mbox{$\mathrm{\Lambda}$-VerToPrimVer}). Here, the off-vertex cut iv) is the main condition 
responsible for the extraction of a $\mathrm{\Lambda}$ signal.
Starting with the moderate conditions as used in the previous high-statistics analysis of the 
$\mathrm{\Lambda}$ phase-space distribution and polarization \cite{hades_lambda_pNb}, a clear 
$\mathrm{\Lambda}$ signal could be separated from the combinatorial background in the p-$\pi^-$ 
invariant-mass distribution. 
While in that analysis a signal-to-background ratio in the order of unity was sufficient, 
for the present $\mathrm{\Xi^-}$ search we start with a higher $\mathrm{\Lambda}$ purity ($>$85\,\%,  
cp. \cite{hades_Xi_ArKCl}). Hence, with the stronger cuts and the requirement  
of an additional $\pi^-$ meson, the number of reconstructed $\mathrm{\Lambda}$ 
hyperons decreases from about 1.1 million to 300,000. 
(No event containing clearly more than one $\mathrm{\Lambda}$ was found.) 
Taking this still high-statistics $\mathrm{\Lambda}$ sample, we
started the $\mathrm{\Xi^-}$ investigation by combining - for each event containing a 
$\mathrm{\Lambda}$ candidate (selected by a $\pm2\sigma$ window around the $\mathrm{\Lambda}$ 
peak) - the $\mathrm{\Lambda}$ with those $\pi^-$ mesons not already 
contributing to the $\mathrm{\Lambda}$. The result was  
a structureless $\mathrm{\Lambda}$-$\pi^-$ invariant mass distribution.   
Hence, additional conditions were necessary: 
v) a lower limit on the 2nd $\pi^-$ (potential $\mathrm{\Xi^-}$ daughter) 
track distance to the primary vertex (\mbox{$\pi_2$-VecToPrimVer}), 
vi) an upper limit of the distance of the $\mathrm{\Xi^-}$ pointing vector w.r.t. the primary 
vertex (\mbox{$\mathrm{\Xi}$-VecToPrimVer}), 
vii) a maximum value of the minimum track distance of the $\mathrm{\Lambda}$ and the 2nd $\pi^-$ 
(\mbox{$\pi_2$-$\mathrm{\Lambda}$-MinVecDist}), and  
viii) a minimum value of the distance of the $\mathrm{\Xi^-}$ vertex relative to the primary one 
(\mbox{$\mathrm{\Xi}$-VerToPrimVer}).

\begin{figure}[!htb]
\begin{center}
\includegraphics[width=1.2\linewidth,viewport=0 0 680 550]{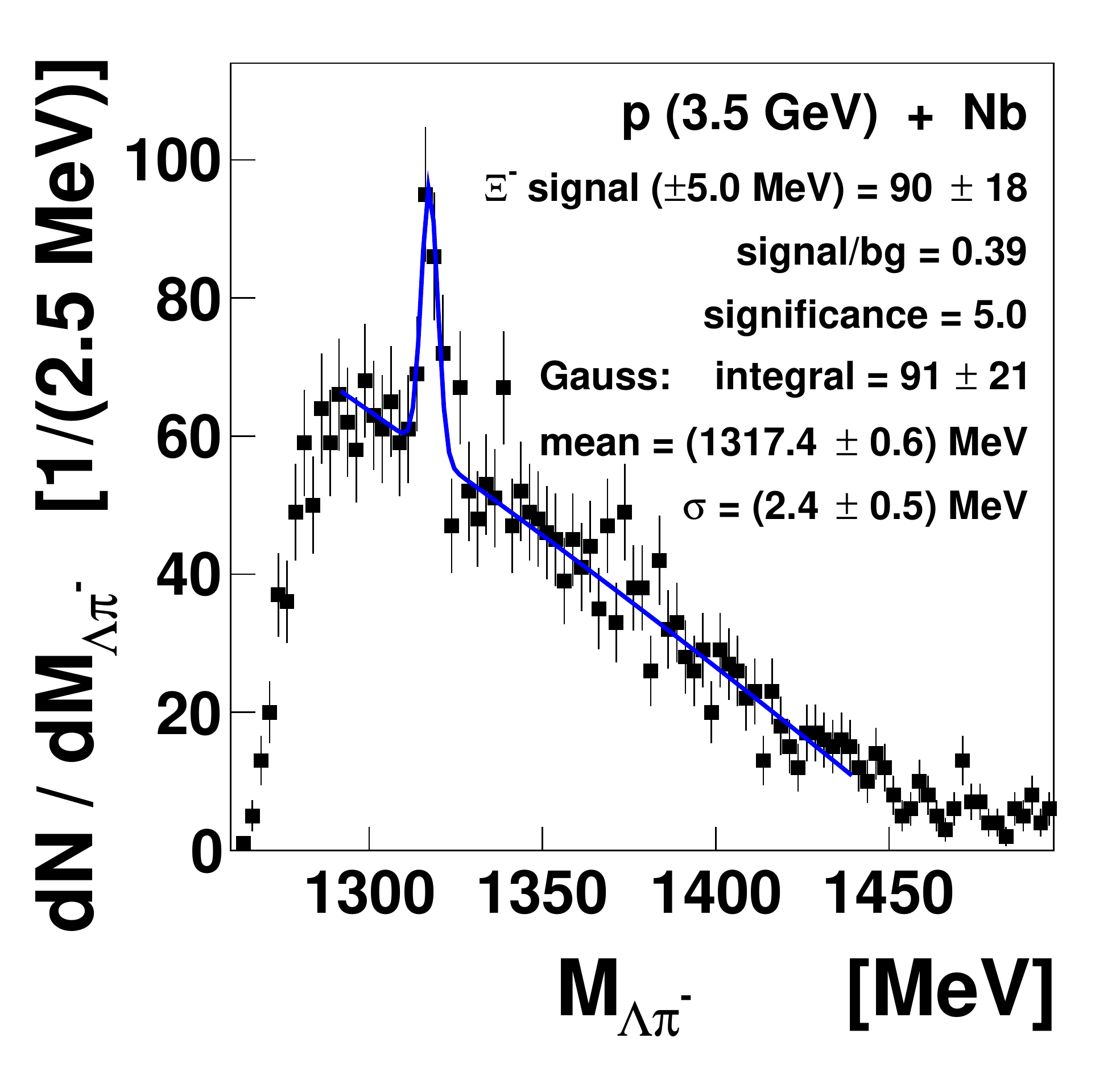}
\caption[]{The experimental $\mathrm{\Lambda}-\pi^-$ invariant-mass distribution. The error bars 
show the statistical errors. The curve represents a combination of a Gaussian and a polynomial 
function used to fit the data. 
\label{xi_mass_exp}}
\end{center}
\end{figure}

Starting with the cut settings used in our previous analysis of deep-subthreshold 
$\mathrm{\Xi^-}$ production in collisions of Ar\,+\,KCl at 1.76$A$~GeV \cite{hades_Xi_ArKCl} and 
optimizing further for the present experiment which exhibits different multiplicities and phase-space 
distributions of the involved particles, we find a significant narrow peak structure in the 
$\mathrm{\Lambda}$-$\pi^-$ invariant-mass distribution displayed in Fig.\,\ref{xi_mass_exp}.
(For convenience, we use identical mass and energy units.)   
The position is slightly lower by about 4~MeV than the PDG value of 1321.71~MeV \cite{PDG2014}, 
very probably due to a minor systematics uncertainty of the momentum calibration for 
charged particles in the inhomogeneous field of the toroidal magnet which leads to a slight ($<$0.4\,\%)  
phase-space dependence of the mass of the reconstructed weakly decaying mother particle as observed 
already for K$^0$ and $\mathrm{\Lambda}$ reconstruction \cite{hades_pp_pNb_K0,hades_lambda_pNb}.  
The width of the present peak, however, is well in agreement with the results of 
GEANT \cite{GEANT} simulations of about 2-3~MeV for $\mathrm{\Lambda}$ and $\mathrm{\Xi^-}$ hyperons. 
More importantly, also the cut dependences of the yield of the potential $\mathrm{\Xi^-}$   
match well those found from GEANT simulations (see below). Thus, we attribute the signal to the 
decay of the $\mathrm{\Xi^-}$ baryon. 

The full curve in Fig.\,\ref{xi_mass_exp} 
shows the result of a fit to the data with a model function consisting of a Gaussian function 
for the peak and a polynomial function of 2nd order for the combinatorial background (bg).   
Integration around the peak maximum within a window of $\pm5$~MeV ($\approx \pm 2\sigma$, with    
$\sigma$ being the Gaussian width) we find 
$N_{\mathrm{\Xi^-}} = 90 \pm 18$ with the statistical error given. 
The signal-to-background ratio and the significance, $\mathrm{signal/\sqrt{signal+bg}}$, 
amount to 0.39 and 5.0, respectively.  
Note that the raw $\mathrm{\Xi^-}$ yield per LVL1 event of 
$N_{\mathrm{\Xi^-}}/N_{\mathrm{LVL1}}=2.8\times 10^{-8}$ is yet a factor seven smaller than the 
corresponding yield in Ar\,+\,KCl reactions at 1.76$A$~GeV \cite{hades_Xi_ArKCl}. 
%
We studied also the raw phase-space distribution of the $\mathrm{\Xi^-}$ baryons. To that purpose, 
the yield within a window of $\pm5$~MeV 
around the $\mathrm{\Xi^-}$ peak in  Fig.\,\ref{xi_mass_exp} 
was selected, and the combinatorial background below the peak was subtracted with the help of a 
corresponding sideband analysis. The resulting transverse-momentum vs. rapidity distribution  
was found strongly biased by the HADES acceptance, i.e. essentially by the lower and upper polar 
angle limits of 18 and 85 degrees \cite{hades_pion_eta_pNb,hades_lambda_pNb}. 
This finding is confirmed by studies of the detector acceptance of simulated data  
and found to be rather independent of the input phase-space 
distributions. The mean value and the r.m.s. width of the experimental rapidity 
distribution amount to 0.54 and 0.16, respectively. The corresponding values of the 
transverse-momentum distribution are 0.52~GeV$/c$ and 0.17~GeV$/c$, respectively.

Corrections for the finite acceptance and reconstruction efficiency were deduced from 
simulations. Thermo-statistically distributed $\mathrm{\Xi^-}$ baryons, 
characterized by a temperature parameter $T$,  
were generated with the event generator Pluto \cite{Pluto}. 
Since the phase-space distribution of the $\mathrm{\Xi^-}$ is not known, 
the experimental $\mathrm{\Lambda}$ phase-space distribution (found to be strongly influenced by 
hyperon-nucleon collisions \cite{hades_lambda_pNb}) served as 
benchmark for the $\mathrm{\Xi^-}$ hyperon. Consequently, in Pluto we allowed for  
longitudinally shifted and elongated $\mathrm{\Xi^-}$ phase-space distributions. For this purpose, 
two longitudinal shape parameter, 
i.e. the mean, $\langle y \rangle=0.3$, and the width, $\sigma_y=0.57$, following 
from a Gaussian fit to the $\mathrm{\Lambda}$ rapidity distribution \cite{hades_lambda_pNb}, are 
introduced.  We investigated the $\mathrm{\Xi^-}$ geometrical acceptance for a broad range of 
transverse and longitudinal shape parameters, i.e. $T=50, 65, 80, 95$\,MeV 
(cf. ref.~\cite{hades_lambda_pNb}), $\langle y \rangle=0, 0.3, 0.6$. 
With the given parameters, we determined, with Pluto, the average HADES acceptance 
for the $\mathrm{\Xi^-}$ hyperon (including the branching ratio of 64\,\% for the decay of 
its daughter, $\mathrm{\Lambda \rightarrow p \pi^-}$) and its variation 
within the parameter ranges. Thus, we estimated a systematic error of 
about $\pm25\,\%$ around the average, purely geometric, $\mathrm{\Xi^-}$ acceptance of 
$\epsilon_{\mathrm{acc},\,sym} = 6.4 \cdot10^{-2}$ for the above given phase-space parameters. 
The same Pluto data are processed through GEANT, modeling the detector response. 
The GEANT data were embedded into 
real experimental data and then processed through the full analysis chain (using the same 
topological cuts i) ... viii) as applied to the experimental data).
The mean value and the r.m.s. width of the resulting HADES-filtered $\mathrm{\Xi^-}$ rapidity 
distribution amount to 0.60 and 0.16, respectively, quite similar to the experimental values.
Relating the output to the corresponding input, 
the total $\mathrm{\Xi^-}$ acceptance $\times$ reconstruction efficiency was estimated to   
$\epsilon_{\mathrm{eff}}= (8.49 \pm 0.24) \cdot 10^{-5}$. 
As in the $\mathrm{\Lambda}$ hyperon analysis \cite{hades_lambda_pNb} we correct for the 
LVL1 trigger bias w.r.t. minimum-bias events, 
$F_{\mathrm{LVL1}}=N_{\mathrm{min \, bias}}/N_{\mathrm{LVL1}}=1.53\pm0.02$, and 
for empty-track events due to non-target interactions, $F_{\mathrm{MT}}=0.17$. 
Finally, we note that the experimental $\mathrm{\Lambda}$ rapidity distribution 
\cite{hades_lambda_pNb} does not appear perfectly symmetric. Also, transport model calculations 
(cf. Fig.\,6 of ref.~\cite{hades_lambda_pNb}) rather predict a faster yield decrease 
in the backward hemisphere than is expected from the rapidity-symmetric Pluto distribution. 
Provided that the $\mathrm{\Xi^-}$ hyperon exhibits a similar asymmetric rapidity distribution as 
the $\mathrm{\Lambda}$, we have to correct the $\mathrm{\Xi^-}$ yield for this difference. We do that 
by calculating the acceptance ratio of 
an asymmetric rapidity distribution to the symmetric one. The asymmetric distribution is modelled 
by a function consisting of two Gaussian distributions. The first Gaussian is the above one describing 
the $\mathrm{\Lambda}$ rapidity density in the region where experimental data points are available and 
the second Gaussian is a more narrow one ($\sigma_y=0.13$, $\langle y \rangle=0.07$) 
following the transport model predictions at 
lower rapidities and joining up to the first Gaussian at $y\sim0.06$. The ratio of the acceptances for 
asymmetric and symmetric rapidity distributions amounts to  
$F_{\mathrm{asy}}=\epsilon_{\mathrm{acc,\,asy}}/\epsilon_{\mathrm{acc,\,sym}}=1.32 \pm 0.02$. 
Assuming that the total $\mathrm{\Xi^-}$ acceptance and reconstruction efficiency can be factorized 
into an acceptance part and and a pure reconstruction-efficiency part which itself does not vary 
within the, rather limited, acceptance, this factor is used to correct $\epsilon_{\mathrm{eff}}$.

With all the necessary quantities and correction factors at hand,   
we calculate the $\mathrm{\Xi^-}$ production probability to 
\begin{equation}
\begin{split}
P_{\mathrm{\Xi^-}} & =\frac{N_{\mathrm{\Xi^-}}}
        { (1-F_{\mathrm{MT}})\,F_{\mathrm{LVL1}}\,N_{\mathrm{LVL1}}\,F_{\mathrm{asy}}\,\epsilon_{\mathrm{eff}}} \\
          & =(2.0\pm0.4\,\mathrm{(stat)}\pm 0.3\,\mathrm{(norm)}\pm 0.6\,\mathrm{(syst)})\times10^{-4},
\label{xi_prob}
\end{split}
\end{equation}
where the statistical, absolute normalization, and systematic errors are given. 

\begin{figure}[!htb]
\begin{center}
\includegraphics[width=1.2\linewidth,viewport=0 0 680 540]{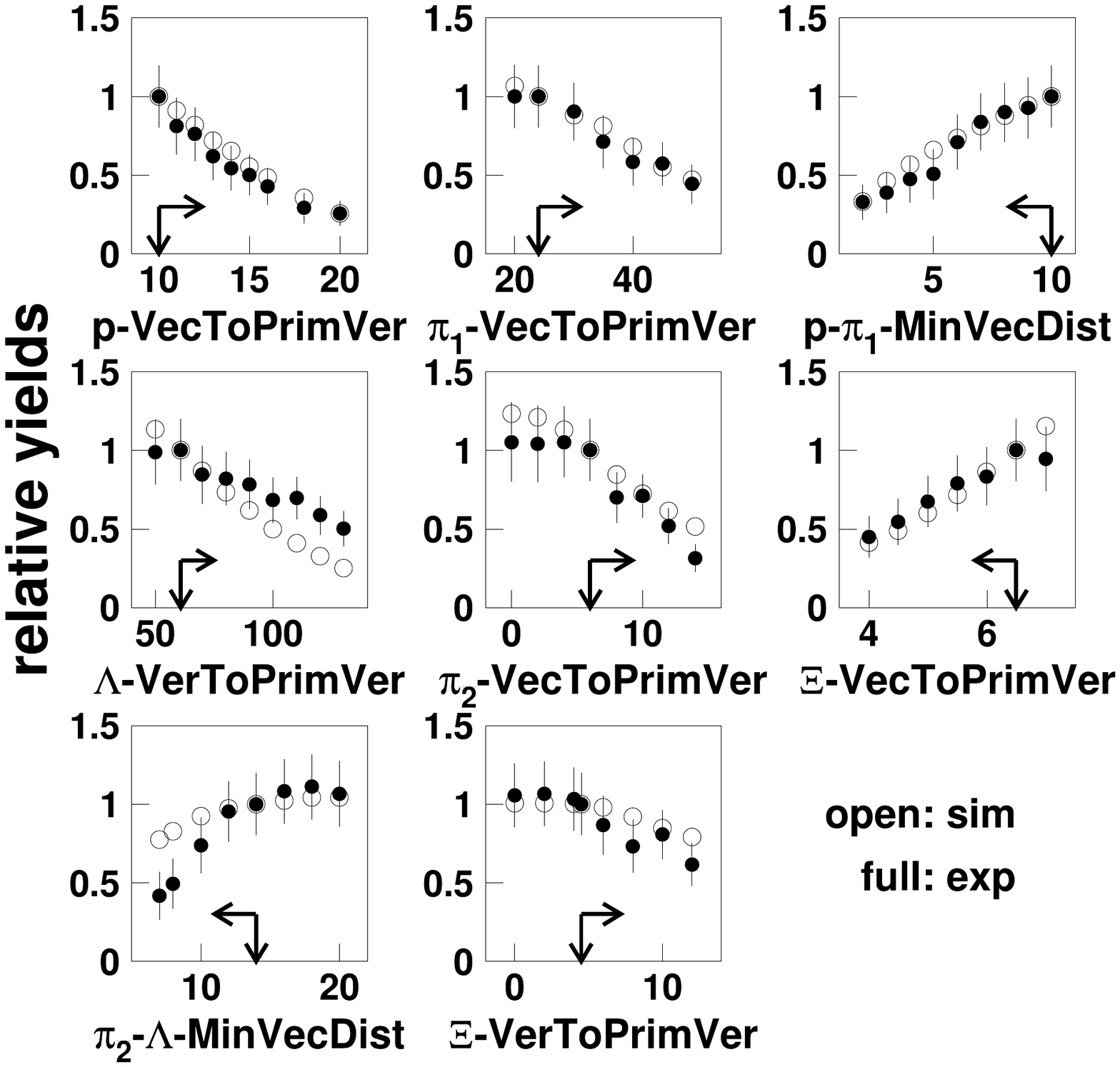}
\caption[]{Relative $\mathrm{\Xi^-}$ yield as a function of the cut value of various 
$\mathrm{\Lambda}$ and $\mathrm{\Xi^-}$ geometrical distances (see text, abscissa units are mm). The 
full (open) circles display the experimental (simulation) data. The vertical and horizontal arrows 
indicate the chosen cut values and the region of accepted distances, respectively.
\label{xi_eff_vs_cuts}}
\end{center}
\end{figure}

The dependences of the $\mathrm{\Xi^-}$ yield on the cut values of the various geometrical 
quantities are displayed in Fig.\,\ref {xi_eff_vs_cuts}. Due to the limited $\mathrm{\Xi^-}$ 
statistics, only one cut could be varied while all the others are kept fixed to the optimum values 
(indicated by arrows) yielding the most significant signal. 
The dependences of experimental data (full circles) 
and GEANT simulations (open circles) are found to be in good agreement. 

Taking the $\mathrm{\Lambda}$ production probability of 0.017 per minimum-bias event 
as estimated in ref.~\cite{hades_lambda_pNb}, the ratio of $\mathrm{\Xi^-}$  
and $\mathrm{\Lambda}$ production yields can be determined. Such a ratio, when derived from the 
same data analysis, has the advantage that systematic errors (e.g. the uncertainty of the absolute 
normalization) cancel to some extent. The ratio amounts to   
\begin{equation}
\frac{P_{\mathrm{\Xi^-}}}
 {P_{\mathrm{\Lambda + \Sigma^0}}}=(1.2\pm0.3\,\mathrm{(stat)}\pm 0.4\,\mathrm{(syst)})\times10^{-2}.
\label{xi_lambda_ratio}
\end{equation}
Here, the statistical error is dominated by the 20\,\% error of the $\mathrm{\Xi^-}$ signal, 
while the systematic error is governed by the stability of the signal against cut variations and  
by the range of the parameters entering the simulation.  

\begin{figure}[!htb]
\begin{center}
\includegraphics[width=1.2\linewidth,viewport=0 0 680 570]{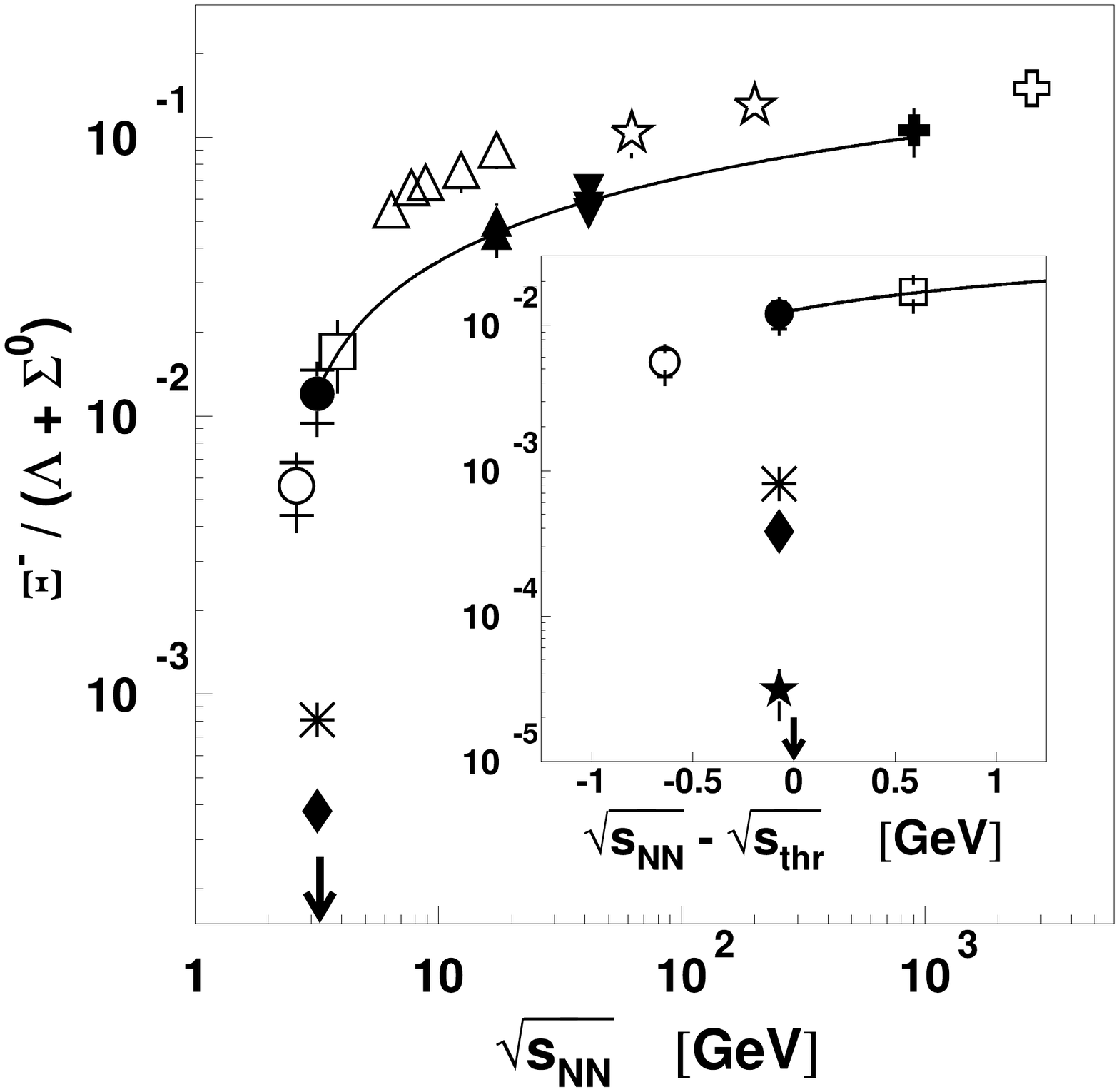}
\caption[]{The yield ratio $\mathrm{\Xi^-/(\Lambda+\Sigma^0)}$ as a function of $\sqrt{s_{NN}}$ or 
$\sqrt{s_{NN}} - \sqrt{s_{thr}}$ (inset). The arrows indicate the threshold in free NN collisions. 
The open symbols represent data for symmetric heavy-ion 
collisions measured at LHC \cite{ALICE_Xi_PbPb_2p76TeV,ALICE_Lambda_PbPb_2p76TeV} (cross), 
RHIC \cite{STAR11,STAR07} (stars), SPS \cite{NA57_04,NA49_08} (triangles), AGS \cite{E895_04} 
(square), and SIS18 \cite{hades_Xi_ArKCl} (circle). The filled cross depicts p\,+\,p collisions 
at LHC \cite{ALICE_pp_900GeV}, while the downward and upward pointing filled triangles are for p\,+\,$A$ 
reactions at DESY \cite{Agari06} and SPS \cite{NA57_06}, respectively. The filled circle 
shows the present ratio (\ref{xi_lambda_ratio}) for p\,(3.5 GeV)\,+\,Nb reactions (statistical error 
within ticks, systematic error as bar). The full curve 
is a parameterization (see text) of the proton-induced reaction data. 
The asterisk, diamond and filled star display the predictions of the statistical-model package 
THERMUS \cite{THERMUS}, the GiBUU \cite{GiBUU,Weil12}, and the UrQMD \cite{UrQMD1,UrQMD2} transport 
approaches, respectively.  

\label{xi_exc_fct}}
\end{center}
\end{figure}

The deduced ratio (\ref{xi_lambda_ratio}) can be compared with corresponding ratios at higher energies 
\cite{ALICE_Xi_PbPb_2p76TeV,ALICE_Lambda_PbPb_2p76TeV,STAR11,STAR07,NA57_04,NA49_08,E895_04,ALICE_pp_900GeV,Agari06,NA57_06}. 
Figure\,\ref{xi_exc_fct} shows a compilation of such ratios as a function of $\sqrt{s_{NN}}$. 
So far, the lowest energy at which a $\mathrm{\Xi^-/\Lambda}$ ratio is available is 
$\sqrt{s_{NN}}=2.61$~GeV, i.e. 630\,MeV below the threshold in NN collisions. The corresponding 
ratio (open circle) was extracted by HADES from Ar\,+\,KCl reactions at a beam energy of 1.76$A$~GeV 
\cite{hades_Xi_ArKCl}.  A steep decline of the $\mathrm{\Xi^-/\Lambda}$ production ratio 
is observed around threshold, where now a second data point (full circle) 
is available at an excess energy of -70~MeV. This allows for comparisons to model calculations 
(see below). To visualize the energy dependence of the proton-induced data 
(full curve in Fig.\,\ref{xi_exc_fct}),   
we fitted the corresponding ratios with a function $f(x) = C(1-(D/x)^{\mu})^{\nu}$ 
(with $x=\sqrt{s_{NN}}$, $C=0.44$, $D=2.2$\,GeV, $\mu=0.027$, $\nu=0.78$), a simple parameterization 
which may be used to estimate the expected $\mathrm{\Xi^-/\Lambda}$ ratio in energy regions,
where data are not yet available.

The $\mathrm{\Xi^-/(\Lambda+\Sigma^0)}$ ratio has been investigated within a statistical approach. 
We performed a calculation with the package THERMUS \cite{THERMUS}, 
using the mixed-canonical ensemble, where strangeness is exactly conserved, while all 
other quantum numbers are conserved only on average by chemical potentials. 
The optimum input parameters for this calculation (i.e. temperature, $T=(121 \pm 3)$~MeV, 
baryon chemical potential, $\mu_B=(722\pm85)$\,MeV, charge chemical potential, $\mu_Q=(24\pm20)$\,MeV, 
fireball radius, $R=(1.05\pm0.15)$~fm, and radius of strangeness-conserving canonical volume, 
$R_c=(0.8\pm2.1)$~fm) follow from the best fit to the available HADES particle yields 
($\pi^-$, $\pi^0$, $\eta$, $\omega$, K$^0$, $\mathrm{\Lambda}$) in p\,+\,Nb collisions at 3.5~GeV 
\cite{hades_pion_eta_pNb,hades_pp_pNb_K0,hades_lambda_pNb}. We obtained a 
$\mathrm{\Xi^-}$ yield of $1.0\times10^{-5}$ and a $\mathrm{\Xi^-/(\Lambda+\Sigma^0)}$ 
ratio of $8.1\times10^{-4}$ (asterisk in Fig.\,\ref{xi_exc_fct}). 
Both values are significantly lower than the corresponding 
experimental data. 

We also estimated the $\mathrm{\Xi}$ production probability within two different 
transport approaches, both having implemented the aforementioned strangeness-exchange 
channels. The first approach is the UrQMD model 
\cite{UrQMD1,UrQMD2} (version\footnote{http://urqmd.org} 3.4). 
For $\mathrm{\Xi^-}$ hyperons, we derived a yield of $(6.9\pm2.8)\times10^{-7}$ per event  
which is more than two orders of magnitude lower than the experimental yield (\ref{xi_prob})  
and decreases only by a factor of two, if the channels $\mathrm{YY \rightarrow \Xi N}$ 
(with cross sections from \cite{FengLi12}) are deactivated; 
i.e. in the model  
hyperon-hyperon fusion is of minor importance for $\mathrm{\Xi}$ production in proton-nucleus 
reactions at 3.5~GeV. The $\mathrm{\Lambda}$ rapidity distribution, however, was fairly well 
reproduced by UrQMD \cite{hades_lambda_pNb}. The resulting $\mathrm{\Xi^-/(\Lambda+\Sigma^0)}$ 
ratio amounts to $(3.1 \pm 1.2)\times10^{-5}$ (filled star in Fig.\,\ref{xi_exc_fct}). 
The second transport approach we used is the GiBUU model \cite{GiBUU,Weil12} 
(release\footnote{https://gibuu.hepforge.org} 1.6). 
We estimated a $\mathrm{\Xi^-}$ yield of $(6.2\pm0.9)\times10^{-6}$, 
a value being considerably higher than the prediction by the UrQMD model, but still 
significantly lower than the experimental yield (\ref{xi_prob}). Also here, 
the total $\mathrm{\Lambda}$ yield was quite well (up to 90\,\%) 
reproduced \cite{hades_lambda_pNb}. 
The $\mathrm{\Xi^-/(\Lambda+\Sigma^0)}$ ratio amounts to $(3.8\pm0.5)\times10^{-4}$ 
(filled diamond in Fig.\,\ref{xi_exc_fct}). The difference of the results of both transport 
models may origin from different parameterizations of cross sections of elementary processes.  

Summarizing, we investigated the production of the $\mathrm{\Xi^-}$ hyperon in collisions of 
p\,(3.5~GeV)\,+\,Nb. For the first time, subthreshold $\mathrm{\Xi}$ production 
is observed in proton-nucleus interactions. 
Assuming a $\mathrm{\Xi^-}$ phase-space distribution similar to that of $\mathrm{\Lambda}$ hyperons, 
the $\mathrm{\Xi^-}$ yield per event amounts to 
$(2.0\,\pm0.4\,\mathrm{(stat)}\,\pm 0.3\,\mathrm{(norm)}\,\pm 0.6\,\mathrm{(syst)})\times10^{-4}$. 
Taking advantage of a recent investigation of $\mathrm{\Lambda}$ hyperon production and polarization 
in the same collisions system \cite{hades_lambda_pNb}, the $\mathrm{\Xi^-/(\Lambda+\Sigma^0)}$ yield 
ratio of $(1.2\,\pm0.3\,\mathrm{(stat)}\,\pm 0.4\,\mathrm{(syst)})\times10^{-2}$ is derived. 
Corresponding estimates with the statistical-model package THERMUS are significantly lower than 
the experimental data, by more than an order of magnitude. 
The GiBUU transport approach predicts a $\mathrm{\Xi^-}$ yield of    
similar level as that of THERMUS. The UrQMD transport model extremely underestimates 
the present data, i.e. the $\mathrm{\Xi^-}$ yield is an order of magnitude lower than the  
GiBUU result. Both transport codes, however, reproduce the pion and single-strange hadron   
yields fairly well. Hyperon-hyperon scattering processes, 
$\mathrm{YY \rightarrow \Xi N}$, recently accounted for the temporary puzzle 
of deep-subthreshold $\mathrm{\Xi}$ production in heavy-ion reactions, are found to be 
of minor importance for subthreshold $\mathrm{\Xi}$ generation in proton-nucleus collisions. 
Hence, a new $\mathrm{\Xi}$ puzzle appears, now in proton-nucleus collisions. 
Probably, direct subthreshold $\mathrm{\Xi}$ production happens on the tail of the Fermi 
distribution of the nucleons within the nucleus, or some of the target protons and neutrons form 
strongly correlated nucleon pairs, kinematically allowing for either direct $\mathrm{\Xi}$ 
production or the population of an intermediate massive resonance decaying into $\mathrm{\Xi}$ 
baryons. Consequently, more systematic investigations are necessary, 
including nucleon-nucleon and pion-induced reactions, goals already pursued by HADES. 
\\[10bp]
The HADES collaboration acknowledges the support by BMBF grants 05P09CRFTE, 05P12CRGHE, 06FY171, 
06MT238~T5, and 06MT9156~TP5, by HGF VH-NG-330, by DFG EClust 153, by GSI TMKRUE, 
by the Hessian LOEWE initiative through HIC for FAIR (Germany), by EMMI (GSI), 
by grant GA~CR~13-067595 (Czech Rep.), by grant NN202198639 (Poland),
by grant UCY-10.3.11.12 (Cyprus), by CNRS/IN2P3 (France), by INFN (Italy),
and by EU contracts RII3-CT-2005-515876 and HP2 227431.

\end{document}